\documentclass[a4paper,11pt]{article}
\usepackage{pos}

\title{Critical Behaviour in the Single Flavor Planar Thirring Model}

\author*[a,b]{Simon Hands}
\author[c]{Michele Mesiti}
\author[a]{Jude Worthy}

\affiliation[a]{Department of Physics, Swansea University,\\
Singleton Park, Swansea SA2 8PP, U.K.}

\affiliation[b]{Department of Mathematical Sciences, University of Liverpool,\\
  Liverpool L69 3BX, U.K.}

\affiliation[c]{Swansea Academy for Advanced Computing, Swansea  University,\\
Bay Campus, Swansea SA1 8EN, U.K.}

\emailAdd{Simon.Hands@liverpool.ac.uk}

\abstract{We report results of simulations of the $2+1d$ Thirring model
with $N$ fermion flavors,
defined on a lattice using domain wall fermions. This approach is devised to
respect as far as possible the underlying U($2N$) symmetry of the continuum 
model, expected to be recovered in the limit wall separation $L_s\to\infty$. For
$N=1$ there is a symmetry-breaking phase transition associated
with bilinear condensation at strong fermion self-interaction, which is a
plausible location for a quantum critical point. Fits to a
renormalisation group-inspired 
equation of state yield critical exponents distinct from those obtained using
a version of the model defined using staggered fermions.}

\FullConference{%
 The 38th International Symposium on Lattice Field Theory, LATTICE2021
  26th-30th July, 2021
  Zoom/Gather@Massachusetts Institute of Technology
}

\begin{document}
\maketitle

\section{Introduction} 
This presentation concerns the Thirring model in $d=2+1$ spacetime dimensions. The
model is a covariant quantum field theory of self-interacting fermions described
by a Lagrangian density
\begin{equation}
{\cal L}=\bar\psi_i(\partial\!\!\!/\,+m)\psi_i+{g^2\over
2N}(\bar\psi_i\gamma_\mu\psi_i)^2.
\label{eq:action}
\end{equation} 
The index $\mu=0,1,2$ runs over spacetime dimensions and $i=1,\ldots N$ over
autonomous fermion flavors. A reducible spinor representation is
chosen so that the $\gamma$-matrices  are $4\times4$, implying the existence of
{\em two} further matrices $\gamma_3$, $\gamma_5$ which anticommute with the
kinetic operator in (\ref{eq:action}). The interaction, with coupling strength
$g^2$ of dimension -1, is a contact term between
two conserved currents, with the consequence that like charges repel, opposite
charges attract. 

Most applications of this and other theories of ``Flatland'' fermions
occur in the condensed matter physics of various layered systems: nodal fermions
in $d$-wave superconductors; spin liquid phases in Heisenberg
antiferromagnets; surface states of topological insulators; and of course
graphene, where low-energy electronic excitations exhibit linear dispersion
around two inequivalent {\em Dirac points\/} within the first Brillouin Zone. 
The Thirring interaction shares the same symmetries as the
electrostatic interaction between electrons and holes, which is unscreened at
half-filling. For sufficiently large $g^2$, and/or sufficiently small $N$, it is
speculated that the Fock vacuum is unstable with respect to formation of a
particle-hole bilinear condensate
\begin{equation}
\langle\bar\psi\psi\rangle={1\over V}{{\partial\ln{\cal Z}}\over{\partial m}}\not=0,
\end{equation}
with clear congruences to chiral symmetry breaking in QCD, resulting in a
dynamically-generated mass gap at Dirac point. By hypothesis, the resulting
semimetal-insulator transition occuring at $g_c^2(N)$ defines a {\em Quantum
Critical Point\/}, whose universal properties characterise the low energy
excitation spectrum~\cite{Son:2007ja}. If a correlation length diverges here,
a new strongly-interacting quantum field theory may be
defined. 

\section{Symmetries}
\label{sec:sym}

On general grounds we expect the QCP to be characterised by dimensionality, the
nature of the degrees of freedom, and the pattern of symmetry breaking. For
$m=0$ the Lagrangian (\ref{eq:action}) has the following invariances:
\begin{equation}
\psi\mapsto e^{i\alpha}\psi,\;\;\bar\psi\mapsto\bar\psi e^{-i\alpha};\;\;\;
\psi\mapsto e^{\alpha\gamma_3\gamma_5}\psi,\;\;\bar\psi\mapsto\bar\psi
e^{-\alpha\gamma_3\gamma_5};
\label{eq:U1U1}
\end{equation}
\begin{equation}
\psi\mapsto e^{i\alpha\gamma_3}\psi,\;\;\bar\psi\mapsto\bar\psi
e^{i\alpha\gamma_3};\;\;\;
\psi\mapsto e^{i\alpha\gamma_5}\psi,\;\;\bar\psi\mapsto\bar\psi
e^{i\alpha\gamma_5}.
\end{equation}
Together these rotations generate U(2). However, 
only (\ref{eq:U1U1}) remains an invariance once $m,\langle\bar\psi\psi\rangle\not=0$. Bilinear
condensation therefore results in a symmetry breaking 
U($2N)\to$U($N)\otimes$U($N$). Note the mass term $m\bar\psi\psi$ is hermitian
and invariant under a parity inversion $x_\mu\mapsto-x_\mu$. In fact, there are
two other parity-invariant mass terms related by U(2$N$) which are
antihermitian in Euclidean metric:
\begin{equation}
im_3\bar\psi\gamma_3\psi;\;\;\;
im_5\bar\psi\gamma_5\psi.
\label{eq:masses35}
\end{equation}
The ``Haldane'' mass term $m_{35}\bar\psi\gamma_3\gamma_5\psi$ is not
parity-invariant, and hence physically inequivalent.

Why is it so important to capture the global symmetries faithfully? The
following argument is very far from rigorous, but illustrates the point.
In the limit of large $N$, the Thirring model (\ref{eq:action}) is amenable to 
a diagrammatic analysis. The interaction at strong coupling between conserved
fermion currents is mediated by a propagating vector boson, (actually a fermion --
antifermion bound state) with mass $M_V$ given by~\cite{Hands:1994kb}
\begin{equation}
{M_V\over m}=\sqrt{6\pi\over{mg^2}},
\end{equation}
whose masslessness in the limit $g^2\to\infty$ suggests
equivalence to 
QED$_3$, an asymptotically-free theory 
long believed to support a conformal IR fixed point. Since the dimensionless
interaction strength at this fixed point scales $\propto N^{-1}$,
there should be a critical $N_c$ below which, again, the ground state is
unstable with respect to bilinear $\bar\psi\psi$ condensation. Since the UV
limit of the Thirring model and the IR limit of QED$_3$ have the same vector
propagator, it is conceivable the fixed points in these limits coincide, so that
the two model share the same $N_c$. For an asymptotically-free theory like
QED$_3$ there is an argument to constrain $N_c$ based on counting
degrees of freedom in both symmetric and broken
phases~\cite{Appelquist:1999hr}:
\begin{equation}
\#\,\mbox{Goldstone bosons in IR}=2N^2\leq{3\over4}\times\#\,\mbox{fermion degrees of freedom
in UV}=3N, 
\end{equation}
where the factor ${3\over4}$ reproduces the Fermi-Dirac distribution correctly in 3
dimensions. Saturating the inequality yields $N_c\leq{3\over2}$
\footnote{A treatment of QED$_3$ based on an
$F$-theorem which does not assume free-field dynamics predicts
$N_c<4.4$~\cite{Giombi:2015haa}}.
Now, many early attempts to identify $N_c$ with lattice field theory used the
staggered fermion formulation. Away from weak couplng staggered fermions support
a different symmetry-breaking U($N)\otimes$U($N)\to$U($N$), yielding a modified
counting:
\begin{equation}
N^2\leq{3\over4}\times2^d\times N\Rightarrow N_c^{\rm stagg}\leq6,
\end{equation}
In fact, numerical simulations with staggered fermions find $N_c^{\rm
stagg}=3.4(1)$~\cite{Christofi:2007ye}.

\section{Domain Wall Fermions}

The arguments of the previous section highlight the importance of choosing a
lattice regularisation which respects the global symmetries of the target theory
as much as possible. We have chosen domain wall fermions (DWF)
formulated in 2+1+1$d$, which recover U($2N$) symmetry in the limit that the
separation $L_s$ of two domain walls along a fictitious third dimension $x_3$
grows large -- though it is not clear {\em a priori} how ``large'' that means.
Fermion fields in the 2+1$d$ target space are supposedly localised on the walls
and are identified in terms of 2+1+1$d$ fields $\Psi,\bar\Psi$ via
\begin{equation}
\psi(x)=P_-\Psi(x,1)+P_+\Psi(x,L_s);\;\;\;
\bar\psi(x)=\bar\Psi(x,L_s)P_-+\bar\Psi(x,1)P_+.
\end{equation}
with projectors $P_\pm={1\over2}(1\pm\gamma_3)$. 

The basic setup was introduced in the context of quenched QED$_3$ in
Ref.~\cite{Hands:2015qha}, where it was noted that U(2$N$) restoration occurs
most rapidly for condensates corresponding to the mass terms
(\ref{eq:masses35}). Henceforth we quote results for $\langle
i\bar\psi\gamma_3\psi\rangle$, which is formed from 2+1+1$d$ propagators
connected to opposite walls. Further details of both the lattice formulation of the
Thirring model used in this study (the so-called {\em bulk} variant) and the
simulation algorithm can be found in \cite{Hands:2018vrd}.

\section{Results}

\begin{figure}
\centerline{\includegraphics[width=10.0cm]{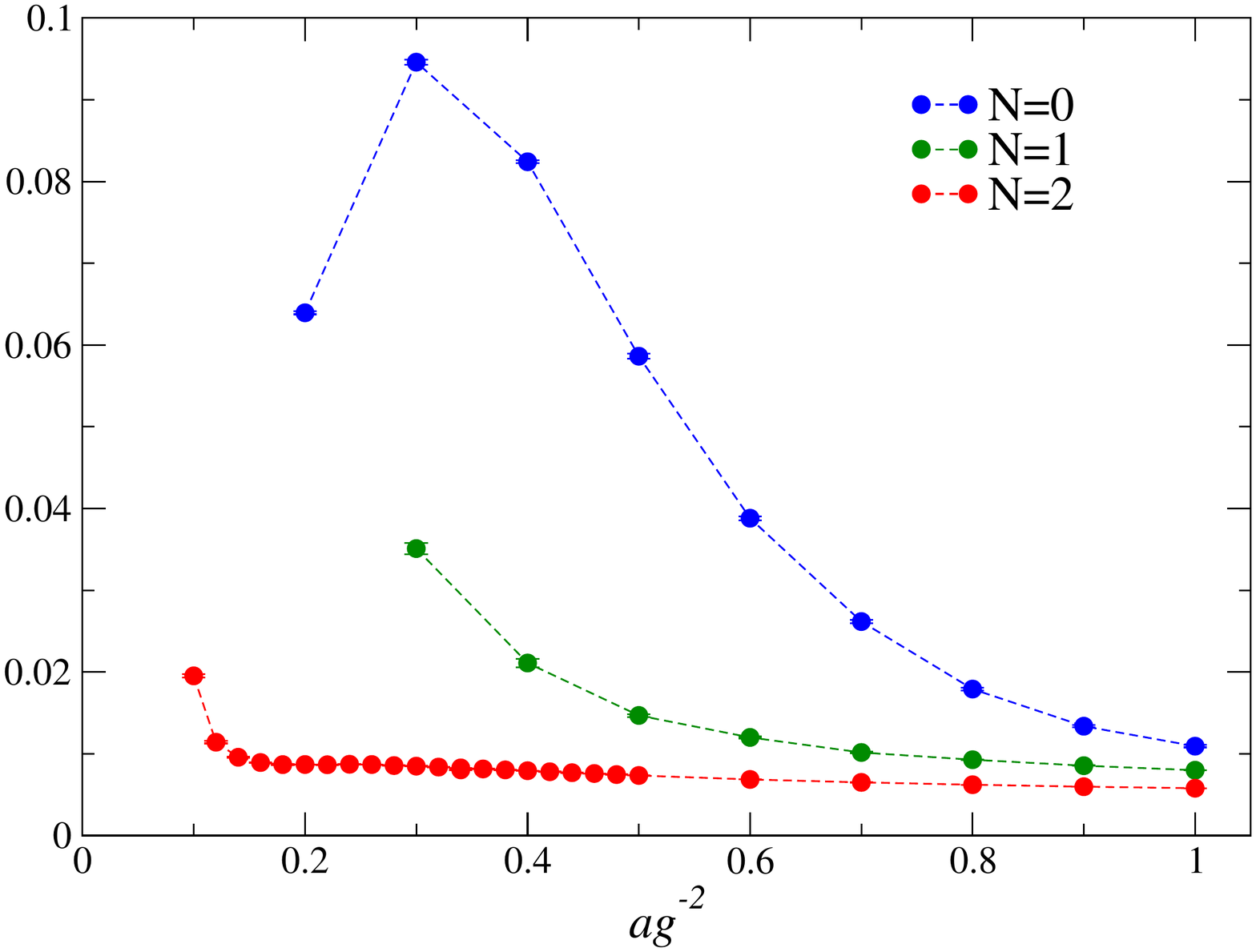}}
\caption{Bilinear condensate $i\langle\bar\psi\gamma_3\psi\rangle$ vs. $ag^{-2}$
on $12^3\times16$ ($16^3\times16$ for $N=0$) with $ma=0.01$.}
\label{fig:N=012}
\end{figure}
\begin{figure}
\centerline{\includegraphics[width=10.0cm]{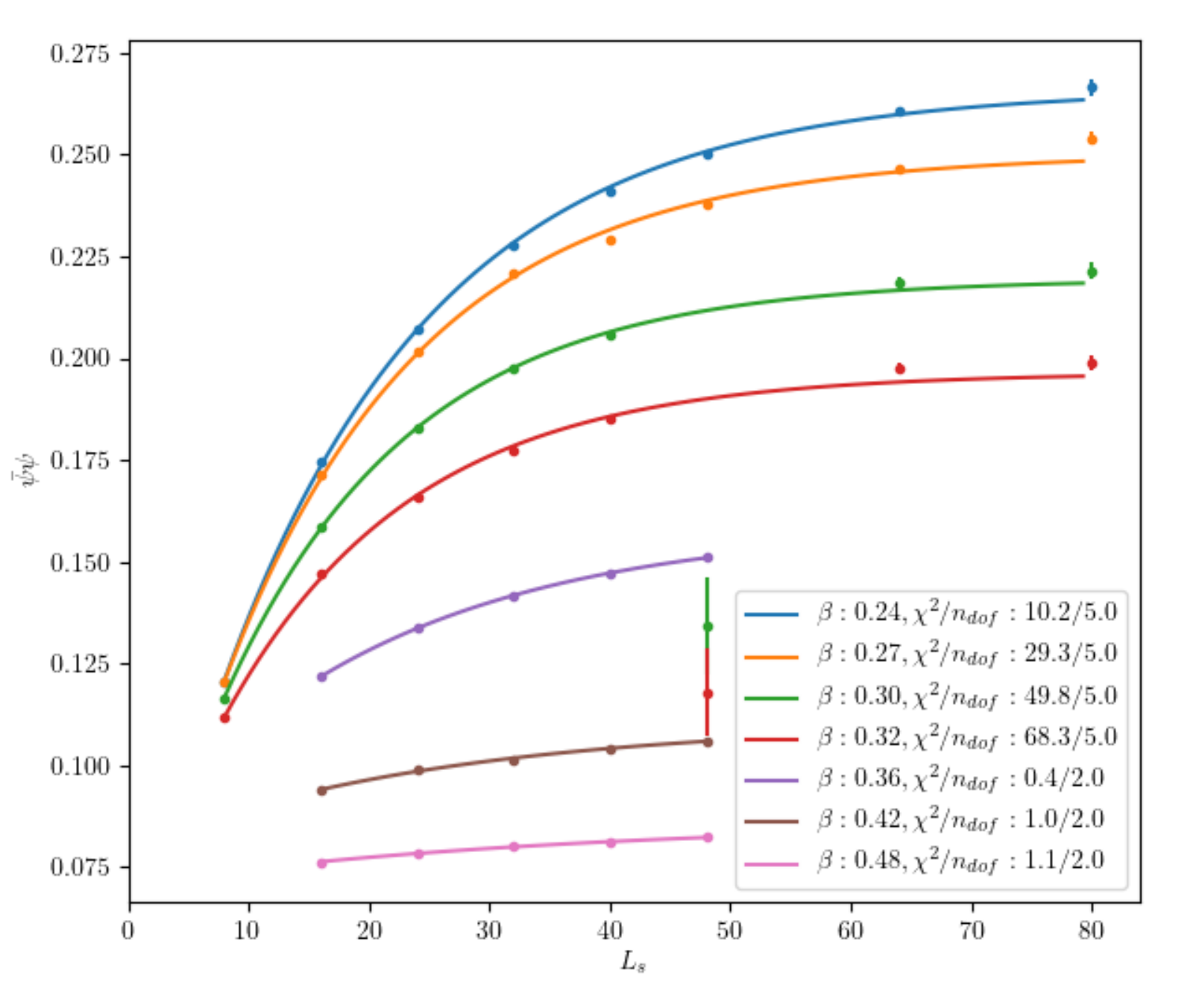}}
\caption{Bilinear condensate $i\langle\bar\psi\gamma_3\psi\rangle$ 
on $16^3\times L_s$ with $ma=0.05$. $\beta\equiv ag^{-2}.$}
\label{fig:pbpextr}
\end{figure}

Fig.~\ref{fig:N=012} presents results for the bilinear condensate as a function
of dimensionless inverse coupling $ag^{-2}$ for exploratory runs with $L_s=16$
~\cite{Hands:2018vrd}, which enable a comparison between quenched
$N=0$, $N=1$ obtained using an RHMC algorithm, and $N=2$ using a much
cheaper HMC algorithm. As might be expected there is a clear hierarchy of
condensation scales
as the coupling gets stronger, confirming our suspicion that this system is very
sensitive to $N$. However, the U(2$N$)-symmetric limit first requires
$L_s\to\infty$. For $N=1$ we have performed systematic studies of system sizes
$12^3,16^3\times L_s$ with $L_s=8,16,\ldots,48$~\cite{Hands:2020itv} and are
currently accumulating data on $16^3$ with $L_s=64,80$. Empirically the extrapolation is
well-described by
\begin{equation}
\langle\bar\psi\psi\rangle_\infty-\langle\bar\psi\psi\rangle_{L_s}=
A(g^2,m)e^{-\Delta(g^2,m)L_s},
\label{eq:extra}
\end{equation}
as shown across a range of couplings in Fig.~\ref{fig:pbpextr} using our latest
data.
The extrapolation (\ref{eq:extra}) is particularly important at larger
couplings, where fortunately it is also easier to fit (difficulties fitting
(\ref{eq:extra}) at weaker
couplings are reflected in the large error bars in Fig.~\ref{fig:eosfit}
below). The decay constant 
$\Delta$ is shown in Fig.~\ref{fig:Delta}.
\begin{figure}
\centerline{\includegraphics[width=10.0cm]{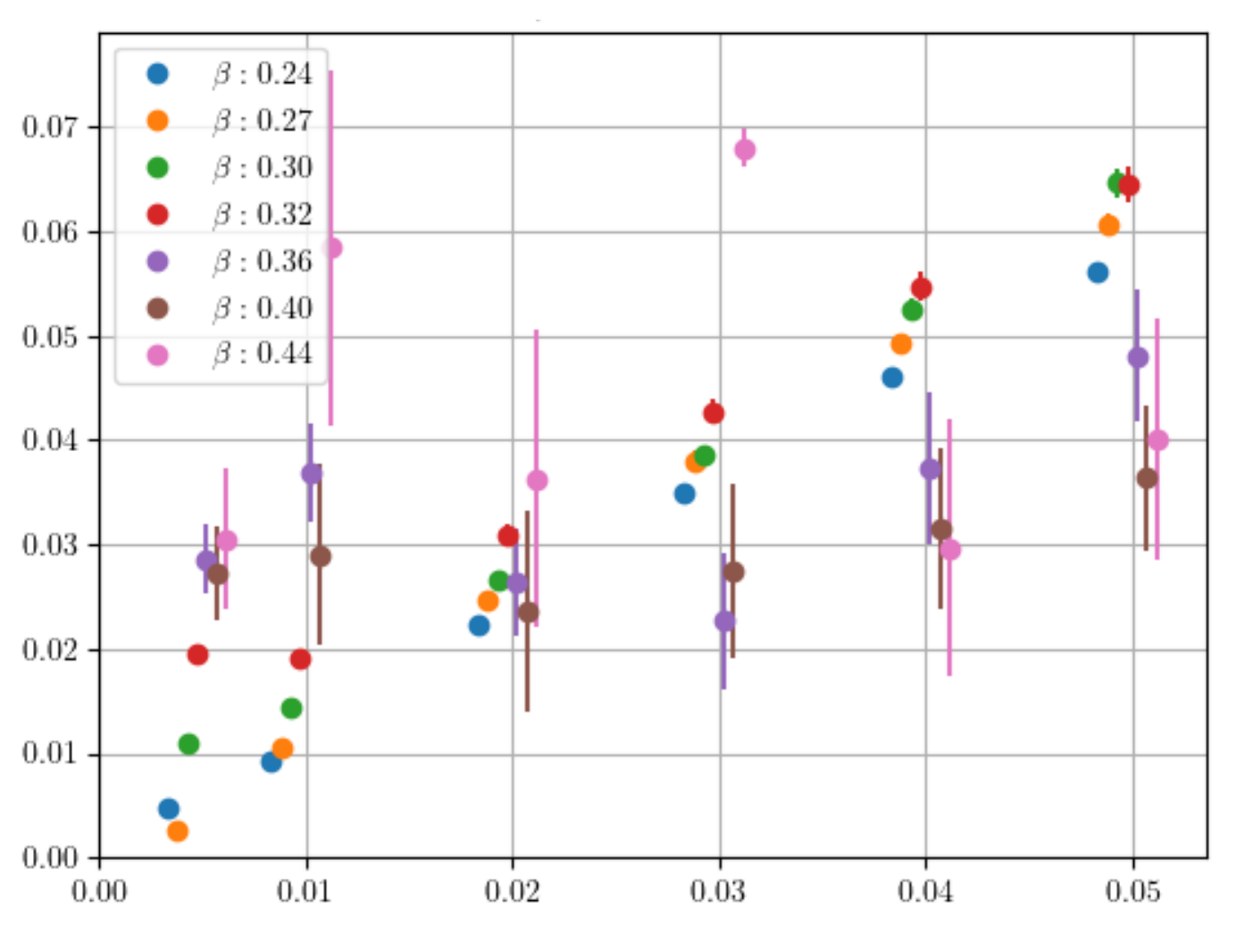}}
\caption{The decay constant $\Delta(g^2,m)$.}
\label{fig:Delta}
\end{figure}
For weak couplings $\Delta$ is approximately $m$-independent 
but once $ag^{-2}\leq0.36$ the behaviour alters to $\Delta\propto m$.
Fig.~\ref{fig:Delta} graphically illustrates the challenge of finding the 
U(2) limit in the strong coupling and massless limits; we are truly
stress-testing DWF. Further aspects of the approach to U(2) symmetry as
$L_s\to\infty$ are discussed
in \cite{Hands:2020itv}.

In order to identify a possible QCP, we need to identify a critical coupling
$g_c^{-2}$ such that in the limit $m\to0$ U(2) symmetry
spontaneously breaks for $g^{-2}<g_c^{-2}$. Since direct estimates of the
$\langle\bar\psi\psi\rangle$ order parameter are not available for $m=0$, our
strategy is to accumulate data for $ma=0.005,0.01,0.02,\ldots,0.05$ and fit the 
whole set to a renormalisation group-inspired equation of
state~\cite{DelDebbio:1997dv}
\begin{equation}
m=A(g^{-2}-g_c^{-2})\langle\bar\psi\psi\rangle^{\delta-1/\beta}+B\langle\bar\psi\psi\rangle^\delta.
\label{eq:eos}
\end{equation}
The latest simulations are aimed at straddling the critical coupling identified
in \cite{Hands:2020itv} with wall separations up to and including $L_s=80$.
\begin{figure}
\centerline{\includegraphics[width=10.0cm]{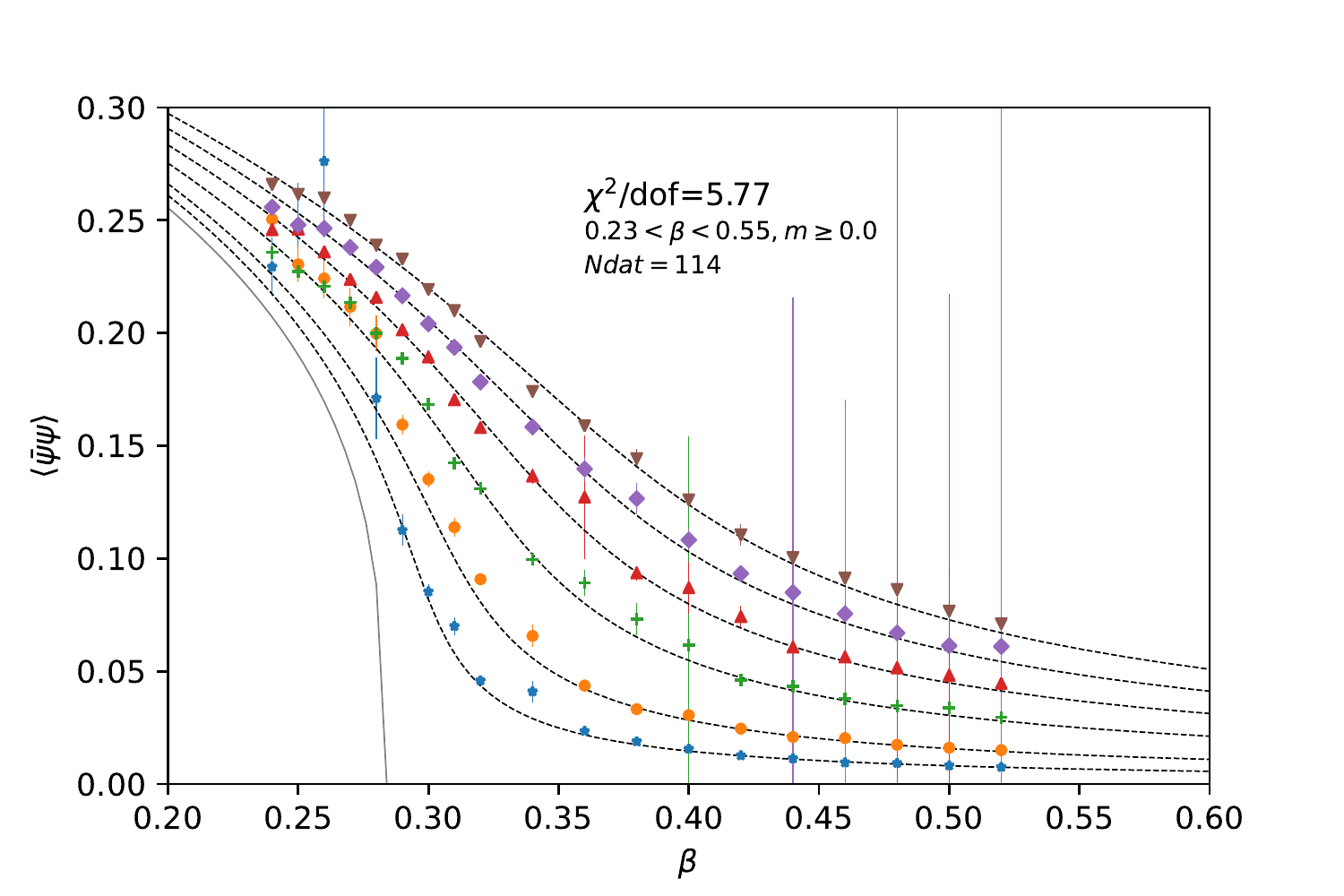}}
\caption{Equation of state fit for $N=1$ on $16^3$ in the $L_s\to\infty$ limit.}
\label{fig:eosfit}
\end{figure}
Our fit to the order parameter data extrapolated to $L_s\to\infty$ is shown in
Fig.~\ref{fig:eosfit}, along with the fitted curve in the $m\to0$ limit.
The critical parameters are 
\begin{eqnarray}
\beta_c\equiv ag_c^{-2}&=&0.283(1);\\
\delta&=&4.17(5);\;\;\;\;\beta=0.320(5),\label{eq:exponents}
\end{eqnarray}
compatible with the fit based on data taken exclusively in the symmetric phase
reported in
\cite{Hands:2020itv}.
Using hyperscaling the critical exponents (\ref{eq:exponents}) can be related
to those more usually extracted from orthodox finite-volume scaling studies:
\begin{equation}
\nu=0.55(1);\;\;\;\;\eta=0.16(1).
\end{equation}

\section{Discussion}

The success of the fit to (\ref{eq:eos}) is strong evidence for 
spontaneous U(2) symmetry breaking and the existence of a QCP for the $N=1$ Thirring model.
Absence of symmetry breaking for $N=2$~\cite{Hands:2018vrd} as  illustrated in
Fig.~\ref{fig:N=012} leads to the conclusion
\begin{equation}
1<N_c<2.
\end{equation}
By way of contrast, investigations of the Thirring model using $N=1$ staggered fermion
flavors~\cite{DelDebbio:1997dv} reveal distinct critical exponents (see also
\cite{Chandrasekharan:2011mn}):
\begin{equation}
\delta=2.75(9);\;\;\;\beta=0.57(2);\;\;\;\nu=0.71(3);\;\;\;\eta=0.60(4),
\end{equation}
and as already observed in Sec.~\ref{sec:sym}, the critical $3<N_c^{\rm stagg}<4$ for staggered
fermions is considerably larger. It seems plausible that the phase transitions 
probed by numerical simulations of DWF and staggered fermions lie in different
universality classes, and define distinct QCPs. This is perfectly
consistent with the remarks made in Sec.~\ref{sec:sym}. It is possible to
formulate a Thirring model with U($N)\otimes$U($N)$ symmetry using
K\"ahler-Dirac fermions in which spinor and taste degrees of freedom are
entangled, which may correspond to the continuum limit of the
staggered model~\cite{Hands:2021mrg} -- there is no reason to expect
``taste symmetry
restoration'' away from weak coupling. 

Before concluding, we should note that other non-perturbative lattice studies
with explicit U(2$N$) symmetry have been performed using fermions with a kinetic
term employing the SLAC derivative~\cite{Lenz:2019qwu}. The results differ
significantly, in particular $N_c^{\rm SLAC}\approx0.8$ implying there is no QCP
corresponding to a local unitary quantum field theory. Clearly the question of
defining suitable regularisations of QFTs away from weak coupling is delicate:
different schemes may fall in the basin of attraction of different fixed
points.

In future work we plan to switch attention to two-point correlation functions,
focussing on both Goldstone and non-Goldstone bound states, as well as the
fermion propagator iteslf, which in principle enables the extraction of a
further exponent $\eta_\psi$. 

\section*{Acknowledgements}
This work was performed using the Cambridge Service for Data Driven Discovery
(CSD3), part of which is operated by the University of Cambridge Research
Computing on behalf of the STFC DiRAC HPC Facility (www.dirac.ac.uk). The
DiRAC component of CSD3 was funded by BEIS capital funding via STFC capital
grants ST/P002307/1 and ST/R002452/1 and STFC operations grant ST/R00689X/1.
DiRAC is part of the National e-Infrastructure. Additional work utilised the
Sunbird facility of Supercomputing Wales. The work of SJH was supported by
STFC grant ST/L000369/1, of MM in part by the Supercomputing Wales project,
part-funded by the European Regional Development Fund (ERDF) via Welsh
Government, and of JW by an EPSRC studentship.


\begin{thebibliography}{99}

\bibitem{Son:2007ja}
D.T.~Son,
Phys. Rev. B \textbf{75} (2007) no.23, 235423.

\bibitem{Hands:1994kb}
S.~Hands,
Phys. Rev. D \textbf{51} (1995), 5816-5826

\bibitem{Appelquist:1999hr}
T.~Appelquist, A.G.~Cohen and M.~Schmaltz,
Phys. Rev. D \textbf{60} (1999), 045003.

\bibitem{Giombi:2015haa}
S.~Giombi, I.R.~Klebanov and G.~Tarnopolsky,
J. Phys. A \textbf{49} (2016) no.13, 135403.

\bibitem{Christofi:2007ye}
S.~Christofi, S.~Hands and C.~Strouthos,
Phys. Rev. D \textbf{75} (2007), 101701.

\bibitem{Hands:2015qha}
S.~Hands,
JHEP \textbf{09} (2015), 047.

\bibitem{Hands:2018vrd}
S.~Hands,
Phys. Rev. D \textbf{99} (2019) no.3, 034504.

\bibitem{Hands:2020itv}
S.~Hands, M.~Mesiti and J.~Worthy,
Phys. Rev. D \textbf{102} (2020) no.9, 094502.

\bibitem{DelDebbio:1997dv}
L.~Del Debbio \textit{et al.} [UKQCD],
Nucl. Phys. B \textbf{502} (1997), 269-308.

\bibitem{Chandrasekharan:2011mn}
S.~Chandrasekharan and A.~Li,
Phys. Rev. Lett. \textbf{108} (2012), 140404.

\bibitem{Hands:2021mrg}
S.~Hands,
Symmetry \textbf{13} (2021) no.8, 1523.

\bibitem{Lenz:2019qwu}
J.J.~Lenz, B.H.~Wellegehausen and A.~Wipf,
Phys. Rev. D \textbf{100} (2019) no.5, 054501.

\end{thebibliography}
\end{document}